# Synchrony and symmetry-breaking in active flagellar coordination

**Kirsty Y. Wan[1,2]**

1. *Living Systems Institute, University of Exeter, Exeter, UK*
2. *College of Engineering, Mathematics, and Physical Sciences, University of Exeter, Exeter, UK*

## Summary

Living creatures exhibit a remarkable diversity of locomotion mechanisms, evolving structures specialised for interacting with their environment. In the vast majority of cases, locomotor behaviours such as flying, crawling, and running, are orchestrated by nervous systems. Surprisingly, microorganisms can enact analogous movement gaits for swimming using multiple, fast-moving cellular protrusions called cilia and flagella. Here, I demonstrate intermittency, reversible rhythmogenesis, and gait mechanosensitivity in algal flagella, to reveal the active nature of locomotor patterning. In addition to maintaining free-swimming gaits, I show that the algal flagellar apparatus functions as a central pattern generator which encodes the beating of each flagellum in a network in a *distinguishable* manner. The latter provides a novel symmetry-breaking mechanism for cell reorientation. These findings imply that the capacity to generate and coordinate complex locomotor patterns does not require neural circuitry but rather the minimal ingredients are present in simple unicellular organisms.

## 1. The locomotor gaits of organisms

Since the invention of photography, the natural habits of organisms have come under increasing scrutiny. Modern optical technologies have enabled resolution of ever-finer detail, so that we can visualise and track behaviour across scales [1, 2]. Consider for example the gallop of a horse, how are its four feet coordinated? This fast action cannot be resolved by the human eye, so the question remained unanswered until 1878, when (reputedly to settle a bet) the photographer Eadweard Muybridge successfully imaged the gait sequences of a horse [3]. Control of rhythmic limb movements in both vertebrates and invertebrates is determined by neuronal networks termed central pattern generators (CPG), where motor feedback is not required [4]. Indeed, CPGs in *in vitro* preparations can produce activity patterns that are identifiable with the *in vivo* behaviour. For instance, the pteropod *Clione limacine* ("sea butterfly") swims with two dorsal wings, the neuronal pattern for basic synchronous movement ("fictive swimming") can be reproduced in isolated pedal ganglia [5]. A striking example of the genetic basis of gait-control and CPGs pertains to the Icelandic horse, where the ability to pace (legs on the same side of the body moving synchronously) was shown to be associated with a point mutation [6].

At the microscale, organisms interact with their world following very different physical principles. Here inertia is negligible and many species developed motile appendages that are slender and suited to drag-based propulsion through fluids. From these simplest designs, myriad locomotion strategies have evolved. The model microswimmer *E. coli* is peritrichously flagellated -- bundles of rigid (prokaryotic) flagella rotate synchronously in the same sense to elicit forward swimming (a "run"), but lose synchrony and unbundle (a "tumble") when one or more of the motors reverse direction (see [7] and references therein). The latter results in rapid changes in reorientation. In this way, a bacterium controls its propensity for reorientation to bias its swimming towards or away from chemo- attractants or repellents [8]. It has been shown that bundling is largely due to hydrodynamic interactions between the rotating filaments [9], but unbundling is stochastic. In contrast, eukaryotes exhibit more deterministic responses to vectorial cues [10]. Phototactic reorientation by the biflagellate alga *Chlamydomonas reinhardtii* requires bilateral symmetry-breaking in a pair of apparently identical flagella [11], though differences in signal transduction must be present [12, 13, 14]. Eukaryotic flagella and cilia possess a flexible and distributed molecular architecture allowing for many more degrees of freedom than prokaryotic flagella. This versatility, particularly when multiple appendages are attached to the same cell, allows algal flagellates to orchestrate diverse swimming gaits such as the breaststroke, trot, or gallop [15].

*Author for correspondence (k.y.wan2@exeter.ac.uk)

Extensive research has already shown that distinct waveforms and beat frequencies can be produced by the same ciliary structure [16, 17] (also Man *et al*, this Issue), depending on intrinsic and extrinsic forcing [18, 19, 20]. This ability to modulate dynein motor cooperativity to produce distinct beating modes on a single cilium does not however explain the higher-level coordination over a network of such oscillating structures. To achieve the latter, intracellular control mechanisms are implicated (reviewed in [21]). In this article, I present new findings on gait patterning in microalgae, which reveal that single-celled multiflagellates can actively dictate the dynamics and activity of each *individual* flagellum. I further propose that symmetry-breaking processes in the flagellar apparatus are causal to this distinguishability between flagella. The attainment of control specificity of locomotor appendages may be a key innovation in the evolution toward increasingly deterministic movement in eukaryotes.

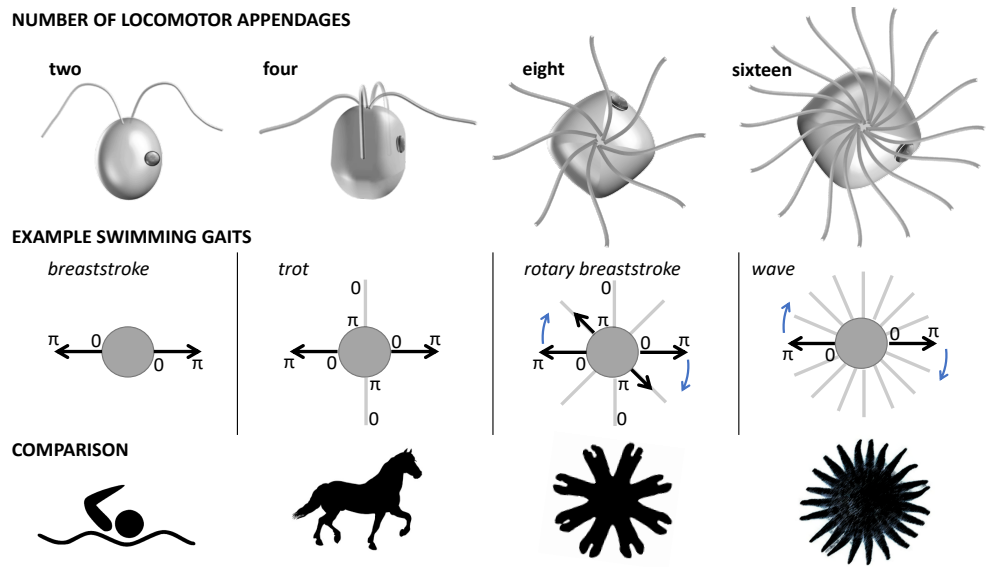

Figure 1. **Locomotor patterning in unicellular algae.** Algae with 2, 4, 8, and 16 flagella are shown with representative swimming gaits. These are characterised by specific phase relationships between the flagella (0 to $2\pi$ for a complete beat cycle). (Different species exhibit different gaits and swimming behaviours even for an identical arrangement of flagella.) Animal models with an equivalent number of locomotor appendages are shown for comparison. These are respectively, a human swimmer, a horse, a jellyfish ephyra (juvenile jellyfish), and the predatory sunflower sea star *Pycnopodia helianthoides* which walks with 16-24 limbs.

## 2. Gait species-specificity

We focus on unicellular algae bearing 2, 4, 8, or 16 flagella, and study the patterns of flagellar actuation and movement. Representative gaits are compared to animals with analogous limb positioning (Figure 1), for which the neuronal basis of movement control has been clearly demonstrated (except in the sea star which is yet to be subject to systematic investigation). In all cases, movement operates in a very different physical and size regime [22, 23] from that encountered by the microeukaryotes from this study. Species-specificity of locomotor behaviours also applies to flagellates, organisms with an identical number and arrangement of flagella can still assume different locomotor gaits [15]. Concurrently, a single species can be capable of multiple gaits associated with significant changes in the flagellar beat pattern. For instance, sudden environmental perturbations elicit so-called shock responses in multiple species, in which flagella reverse their beating direction (but not the direction of wave propagation) to produce transient backward movement or even cessation of swimming [24, 25]. Some species have a stop state [26], in which all flagella become reversibly quiescent. We organise this section by characterising first the gaits used for forward swimming, and then separately, introducing the phenomenon of selective beat activation in free-swimming cells.

### 2.1. Run gaits for forward swimming

The biflagellate *C. reinhardtii* is favoured for biophysical models of swimmers [27]. Cells execute a forward breaststroke gait in which the two flagella are synchronized, interrupted by phase slips. Stochastic switching between this characteristic in-phase breaststroke and a perfectly antiphase (or freestyle) gait was also discovered in a phototaxis mutant [28]. Experimental and theoretical evidence overwhelmingly implicate intracellular coupling as the dominant mechanism for flagellar coupling [29, 15, 30, 31]. Phase-synchrony is not a given however, heterotrophic biflagellates such as *Polytoma uvella* have two flagella of slightly unequal lengths and different beat frequencies, which remain largely asynchronous. With more than two flagella, new gaits become accessible. Depending on flagella configuration and ultrastructure, different interflagellar phase patterns are sustained during steady, forward swimming. Comparison with quadrupedal locomotion is instructive here [32]. Among species with four flagella, a robust *trotting* gait comprising two alternating pairs of synchronous breaststrokes is assumed by two marine *Pyramimonas* species: *P. parkeae* and *P. tychotreta*, whereas *galloping* gaits are consistently displayed by *Tetraselmis sp.* and *Carteria crucifera* (see SM Videos 1-5). Only three extant species are known to have eight flagella [33]. Of these, *Pyramimonas octopus* coordinates its eight flagella into a unique forward propulsion gait which we term the *rotary breaststroke* (see section 4). Identifying equivalence between this and octopedal animal movement is more challenging. The spider possesses obvious bilateral symmetry not present in the algae, though alternating patterns of leg movements are observed [34]. Jellyfish ephyrae use cycles of contraction and relaxation involving simultaneous movement of their eight soft appendages, but move in an inertial regime [35]. Finally, the Arctic species *Pyramimonas cyrtoptera* has 16 flagella [36] which can be actuated metachronously rather than in discrete patterns. This may be the critical number at which the flagella begin to interact hydrodynamically [15]. In this limiting scenario, the wave-like coordination of many closely-separated appendages may help to avoid collisions, and mirrors the locomotion mechanisms adopted by arthropods [37, 38].

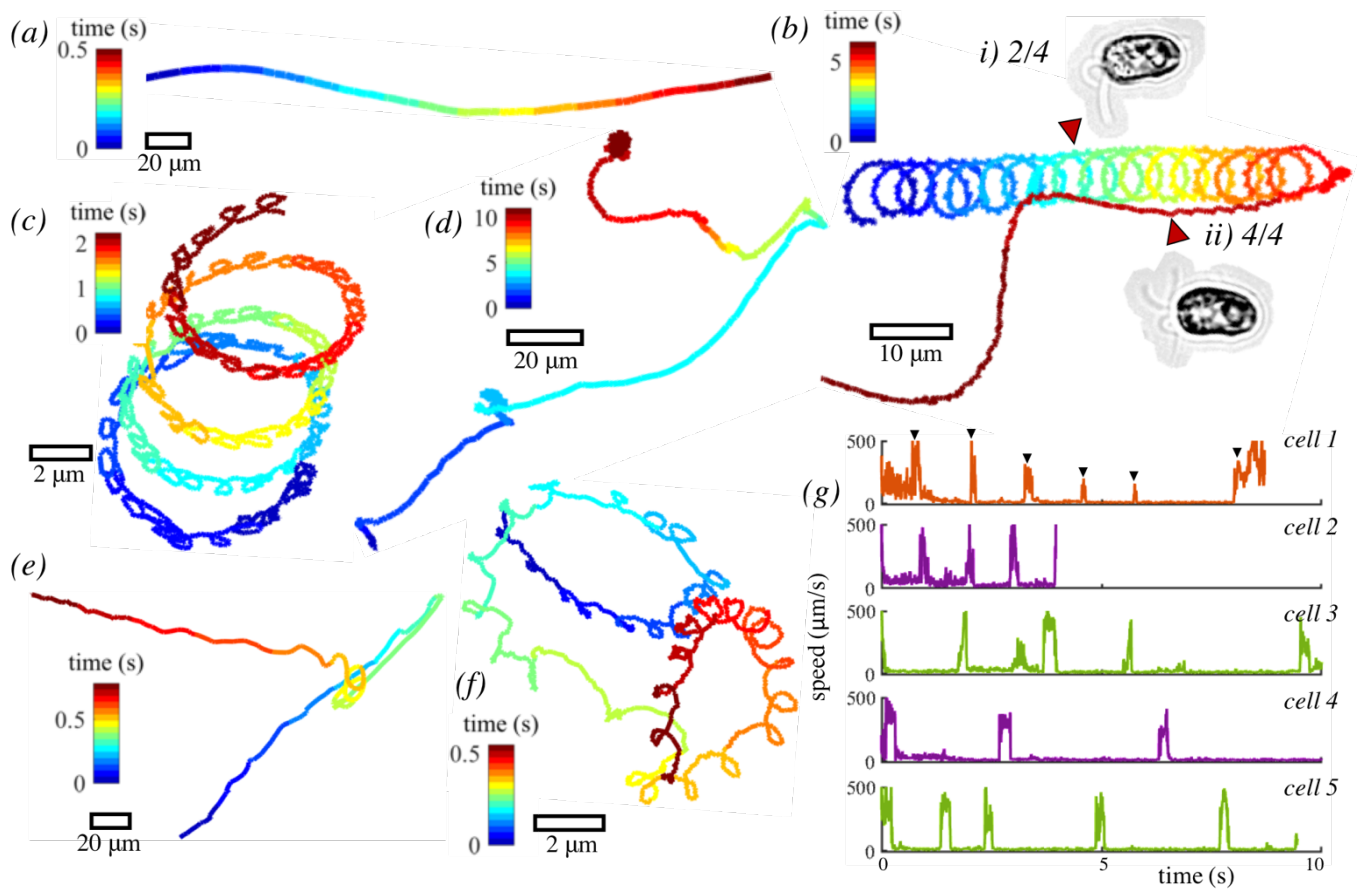

Figure 2: **Active gait reconfiguration and heterogeneity in single-cell swimming trajectories**. Persistent directional swimming (a) is contrasted with asymmetric and chiral trajectories (b-f). Free-swimming cells of the quadriflagellate *Tetraselmis subcordiformis* are shown in (b,c). In (b), a cell transitions from a spinning gait of two flagella (i), to a galloping gait of all four flagella (ii). In (c), beating is restricted to only one of four flagella, resulting in a strongly helical trajectory. A different quadriflagellate *Pyramimonas parkeae* exhibits intermittency: in trajectory (d) a cell alternates between a trot gait of four flagella (active state), and a stop state with no flagella activity (inactive state). In (e), the octoflagellate *Pyramimonas octopus* undergoes reorientation following a shock response (see also [26]). (f) A *P. parkeae* cell presenting a single actively beating flagellum. (g) The swimming speed of *P. parkeae* was measured in a population of individuals, which shows a characteristic activity timescale for "bursting". [SM Videos 1-3]

### 2.2. Gait-switching and partial-activation

In animals, the capacity to produce multiple locomotor modes is critical for survival. Gait often depends on the speed of desired movement. For example, the ghost crab becomes bipedal at the highest speeds [39]. Likewise, we find that unicellular flagellates dynamically reconfigure their motor apparatus to produce different gaits that are coupled to photo- or mechano- sensory pathways. Next, we compare the motion repertoire of different multiflagellate species. Freely-swimming cells were imaged at kilohertz and sub-micrometer resolution, so that flagella motion could be discerned at the same time as the cell body movement. Strong behavioural stereotypy was observed within each species, but heterogeneity across species (Figure 2). Consistently, tracks presenting high curvature loops were found to result from a previously uncharacterised motility mechanism – namely, the partial activation of flagella. For the same cell, beating can be restricted to a subset of flagella with the remaining flagella quiescent. Asymmetries arising from the actively beating

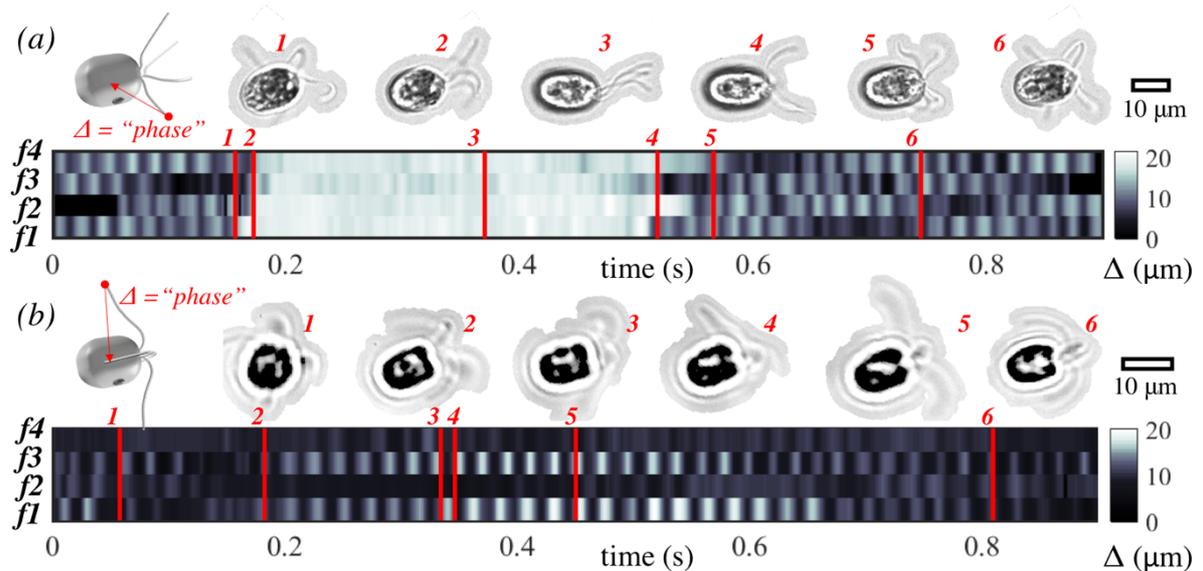

Figure 3: **Quadriflagellate gaits resolved at the single-flagellum level in free-swimming cells.** Displacement from tip to centre of the cell body was used as proxy for flagellar phase for each flagellum, labelled (f1-4). Frames corresponding to specific timepoints are highlighted. Scalebars for cell size reference. (a) *Carteria crucifera* has a galloping gait during forward runs, but exhibits noise-induced shocks involving reorientation of the flagella (time 3). Recovery of the original rhythm (phase-resetting), completed within 4 beat cycles. (b) *Pyramimonas tychotreta* has a canonical trotting gait during forward runs, but exhibits partial activation of flagella. Beating is restricted to the pair f1,3, but suppressed in f2,4. Synchrony of f1,3 drifts from antiphase (times 3,4) to in-phase (time 5) over the course of this recording. [SM Videos 4, 5]

flagellar subgroup produces large-angle turns in the free-swimming trajectory. A previously quiescent flagellum can subsequently become active again, and vice versa (Figure 2b, and SM Video 1). Four different quadriflagellate species showed selective activation of flagella (*P. parkeae, T. suecica, T. subcordiforms,* and *P. tychotreta*). In cells exhibiting motion intermittency such as *P. parkeae* (see Figure 2d), flagella alternate between periods of bursting activity and total quiescence (Figure 2g, SM), analogous to the physiological behaviour of motor neurons.

Next, we resolve gait transients and interflagellar phase dynamics while keeping track of flagellum identity. Automated methods could not follow the flagella movement reliably. Due to the 3D movement and continuous cell axial rotation, features of interest do not always remain in focus. Instead, flagella tip positions were manually annotated in Trackmate [40]. The distance between flagellum tip and cell centroid undergoes oscillations for an actively beating flagellum, but remains constant for a quiescent flagellum. We use this distance as proxy for flagellar phase (Figure 3). The quadriflagellate *C. crucifera* displays stochastic gait-switching between forward swimming and transient shocks, as does *P. octopus* [26]. Shocks involve simultaneous conversion of all four flagella to a hyperactivated undulatory state (Figure 3a). After ~0.3 s, the canonical ciliary beat pattern is recovered. A rhythmic galloping gait resumes after only a few asynchronous beat cycles. The rapidity of phase-resetting responses in these organisms further implies gait control is active [41]. In Figure 3b, we demonstrate partial activation in *P. tychotreta* [42]. Here, full beat cycles are sustained in flagella f1,3, while beating in the diametrically opposite pair f2,4 is suppressed throughout. Small fluctuations detected in the inactive flagella are due to passive movement induced by the beating of nearby flagella. Further examples of quadriflagellates in which beating is dynamically constrained to one, two, or three out of a total possible four flagella, are provided in the SM.

## 3. Intermittency and temporal ordering in a flagellar network

Free-swimming individuals are not amenable to long-time imaging. In order to investigate the long-time statistics of locomotor patterning, experiments were also conducted on micropipette-held (body-fixed) organisms. Single or double micropipette configurations were used to position, manipulate single cells, and to apply localised mechanical perturbations to single flagella. Not all species were suitable for the micropipette technique however - in some cases captured cells prematurely shed their flagella. This is likely related to a stress-induced, calcium-dependent deflagellation response [43, 44]. We identified two quadriflagellate species that were minimally affected by pipette-capture. These are *P. parkeae* and *T. suecica*, which are representative of the two known quadriflagellar arrangements: type I and II in [15] (see also Figure 4). In *P. parkeae* the flagella form a cruciate arrangement, while in *T. suecica* they align into two antiparallel pairs. Both flagellates show partial activation and beat intermittency.

For pipette-fixed individuals, flagellar phases could be extracted automatically from imaging data to characterise the emergence and decay of coordination in the flagellar network. Phase patterns are reminiscent of classical studies of footfall patterns and limbed locomotion in animals [45, 46]. In long-time, high-speed recordings, reversible transitions in locomotor patterning were found in both species. These occurred spontaneously, but could also be induced by mechanical forcing. This touch-dependent, gait-sensitivity is a novel manifestation of a mechanoresponse in motile cilia and flagella. At the moment of capture, contact with the pipette can induce rhythmogenesis in a cell with initially quiescent flagella (Figure 4a, SM Video 6). Similarly, changes in locomotor patterning were induced when fluid was injected manually (~1s pulses) via a second pipette in the vicinity of a quiescent cell (Figure 4b). The first pulse administered at time t1 induced beating in f1,3 only. This biflagellate state continued until the second pulse (time t2), which then resulted in beating in all four flagella (SM Video 7). In both cases, even after cessation of external perturbations, the cells continued to exhibit intermittent activity over tens of seconds. We conclude that spontaneous transitions in behaviour measured during free-swimming (Figure 2) are replicated in pipette-fixed individuals.

Next, we emphasize two aspects of gait reconfigurability in multiflagellates. First, neither activation (e.g. Figure 4c) nor deactivation of beating (e.g. Figure 4d) necessarily occurs simultaneously in all four flagella. However, the shock response (which is an all-or-none response depending on threshold perturbation magnitude [26, 47] is associated with a strong process asymmetry, the forward reaction (activation) is simultaneous but the reverse (deactivation) is sequential - different flagella can cease beating at different times (Figure 4b). Second, after any gait transition, the reestablishment of an expected run gait (whether the trot, or gallop) is not instantaneous, but rapid. The canonical temporal ordering of the flagella is recovered after only a few beat cycles.

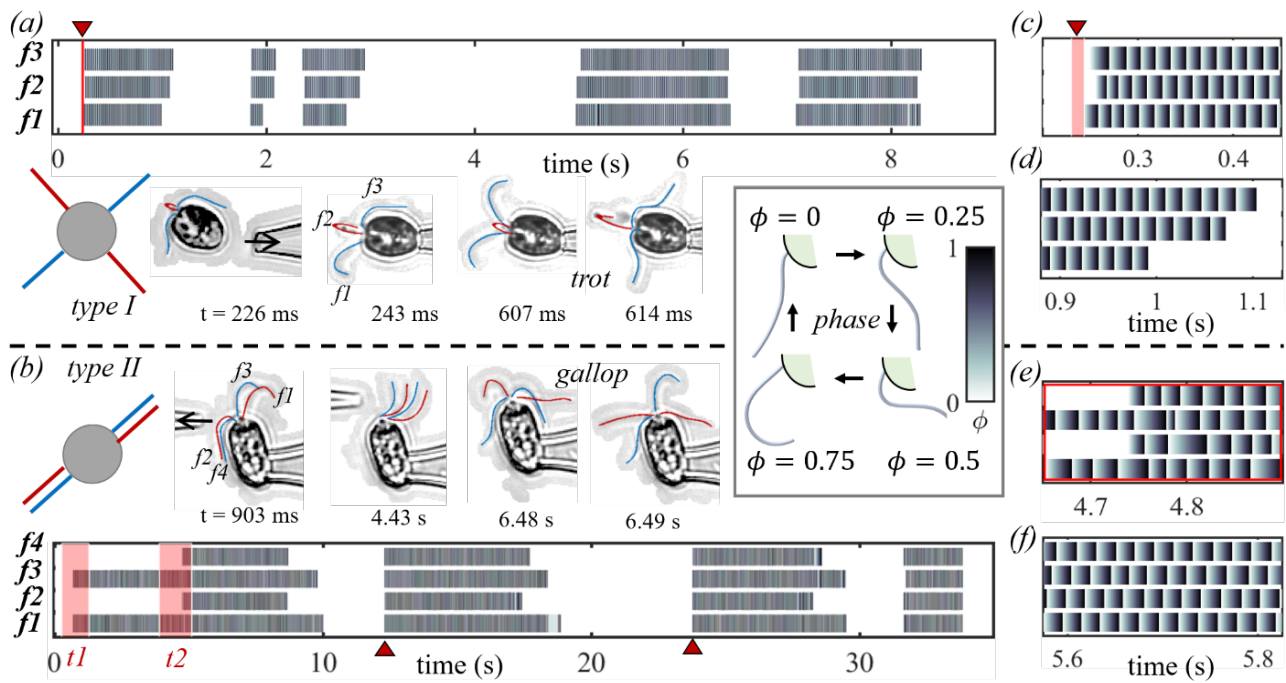

Figure 4: **Reversible rhythmogenesis and gait-mechanosensitivity in quadriflagellates.** (a) The four flagella of *P. parkeae* have a cruciate arrangement (type I, top view). The captured cell has two flagella in the focal plane (f1,3) and two others moving transversely – these are labelled as f2 and are tracked as one (see SM Video 6). (b) The four flagella of *T. suecica* have an anti-parallel arrangement (type II, top view). Mechanical perturbations are introduced at times t1,2, and spontaneous shocks are detected at times indicated by red pointers. (c-f) are zoomed-in plots of (a-b), at time points of interest. At the moment of capture the flagella begin to beat, but not simultaneously; a regular trot gait emerges within 3 beat cycles (c). Spontaneous termination of beating does not occur at the same time in all flagella (d). Induced changes in locomotor patterning (e) are compared with the regular phase patterning (f) observed during the trotting gait. (See SM Video 7). Inset: flagellar phase is defined according to progression through the beat cycle.

## 4. Symmetry-breaking in the flagellar apparatus

In this section we elaborate on the non-identity of flagella in multiflagellate algae. Adapting terminology from gait analyses of limbed animal locomotion, we distinguish between symmetric gaits (e.g. the trot, pronk), and asymmetric gaits (e.g. the gallop, bound). A quadrupedal gait is referred to as being symmetric if footfall patterns of both pairs of feet are evenly-spaced in time, and as asymmetric if the activity of at least one pair is unevenly spaced in time. Different gaits can therefore arise from the same underlying locomotor circuit. The presence of more than two flagella also allows for diverse possibilities for symmetry-breaking even during regular gait patterning (not transitional gaits).

We focus now on the rotary breaststroke of *P. octopus*, which has not been described in any other organism. Gait ordering in multiflagellates is particularly difficult to visualise, due to the highly three-dimensional swimming. Additionally, *P. octopus* rotates slowly clockwise about its own axis when viewed from the anterior end (SM Video 8). All eight flagella are distinguishable simultaneously only when the cell is swimming into or out of the focal plane (Figure 5, SM Video 9). Recordings of such transients were made, from which the extremal positions of each flagellum could be followed in time and tracked manually in ImageJ. As before, a normalised flagellar phase was computed for each flagellum. We find that phase ordering propagates directionally, in the same sense as the about-axis rotation (Figure 5c). (Gaits are similarly directional in the related *P. cyrtoptera*, which swims with 16 flagella.) A pairwise Pearson-correlation matrix was computed for each tracked cell, to quantify the degree of correspondence between the eight flagella (Figure 5d). Strong positive correlation was measured between diametrically opposite flagella (dashed white lines are guides), which tend to move synchronously to produce a noisy sequence of intercalated breaststrokes (first 1:5, then 2:6, and so on).

Given the directional nature of this swimming gait, asymmetries in developmental patterning must be present. This is corroborated by the only ultrastructural study available on this species [48]. The flagella emerge from separate basal bodies (centrioles during cell division), which are localised to the anterior of the organism (Figure 5a). Electron microscopy sections reveal a complex basal architecture lacking in rotational symmetry. Individual basal bodies are positioned uniquely according to generational age [48]. A contractile network of thick and thin fibres (some are striated) connect specific numbered microtubule triplets to provide intracellular coupling between flagella [15]. No obvious structural or morphological differences have been reported in the flagellar axonemes. Consequently, it is not possible from light microscopy data to prescribe the identity of the individual flagella in accordance with the numbering system devised from TEM images. Flagella numbering is thus distinguishable only *up to circular permutation*. Nonetheless, robust in-phase synchrony during free-swimming pertains to diametrically opposite flagella (positive phase correlation, Figure 5d). This is in good agreement with the presence of direct physical coupling between distinguished pairs of flagella. Particularly, the principal pair b1 & b5 is connected by a thick striated structure known as the synistosome, which is functionally related to the distal fibre in *C. reinhardtii* [49]. The lack of correlation between certain basal body pairs (Figure 4d) will be investigated further in future studies to determine if this is related to the detailed topology of the fibre network (e.g. b2 and 8 appear unconnected).

Finally, we find that in ~5% of cases, free-swimming cells exhibit a distinguished flagellum which is held extended in front of the cell body (SM Video 10). Figure 5 compares this "search gait" with the canonical forward swimming gait of this organism. The beat pattern in this single flagellum is undulatory (sperm-like), in contrast to the lateral (ciliary) stroke used by the other seven flagella. These observations further highlight the capacity for producing heterodynamic behaviour in a network of outwardly-identical flagella.

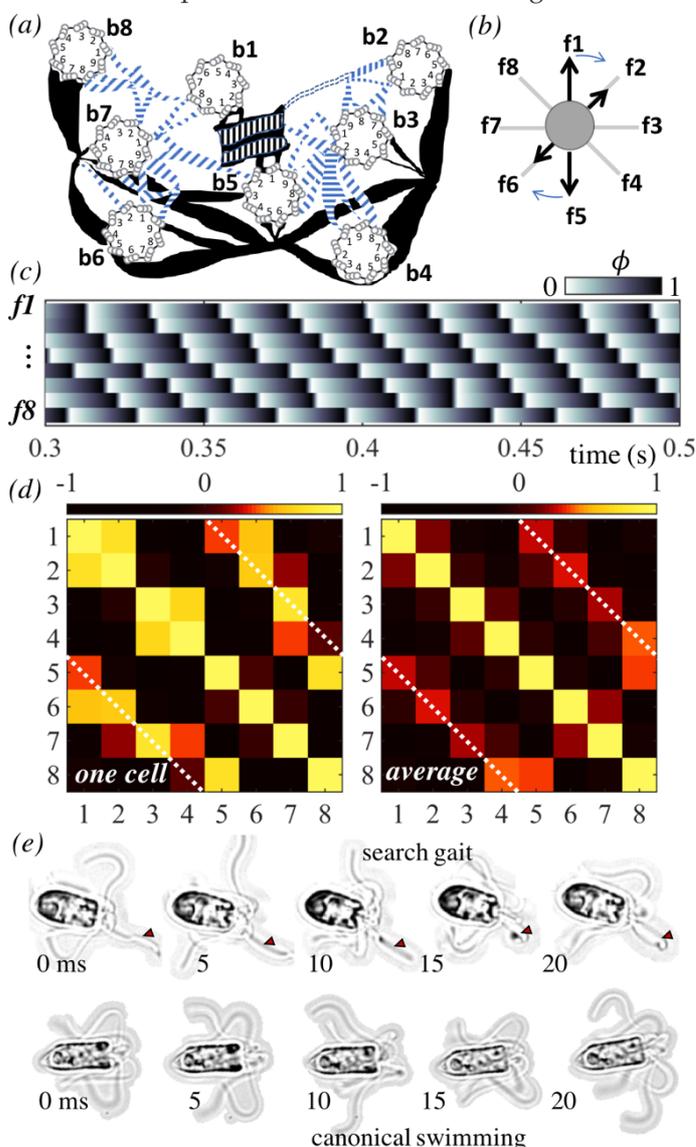

Figure 5: *P. octopus* exhibits symmetry-breaking in both structure and movement. a) Schematic (redrawn from [48]) of the basal apparatus comprising eight basal bodies (b1-b8) whence emanate the flagella (f1-f8). Thick and thin filaments couple the flagella at specific locations to produce the octoflagellate's unique rotary breaststroke (b). Stereotypical phase dynamics for the numbered flagella are displayed in (c). Cross-correlation matrices for one cell, and group average (n = 8), both show strong synchrony between diametrically-opposite flagella (d). (e) A small number of cells adopt a "search" gait, in which a single flagellum (marked by the red arrow) is held extended in front of the cell. (SM Video 10). Canonical swimming is also shown for comparison.

## 5. Discussions and Outlook

The present article is devoted to the gait dynamics of free-living unicellular algal flagellates, advancing the state-the-art from single-cell to single-flagellum resolution. Performing a comparative study, I showed that gait coordination in algal flagellates is not only intracellular but also active. The same network of flagella can undergo dynamic reconfiguration over fast timescales to generate different locomotor patterns. Multiple species share an ability to reversibly activate beating in a subset of flagella, while suppressing the activity of the remainder. This is an extreme instantiation of symmetry-breaking in the algal flagella apparatus. Individually, a eukaryotic flagellum or cilium can be in an oscillatory or non-oscillatory state, beating can be modulated by extrinsic factors including mechanosensitivity [50, 26] (and Bezares Calderon *et al*, this Issue), but how can drastically distinct beat patterns be produced in different flagella *of the same cell*? We suggest two sources of this apparent control specificity in coupled flagella: i) physical asymmetries in the basal apparatus, and ii) differences in signal transduction. The two contributions are not mutually exclusive.

In some species, physical asymmetries are obvious. Heterokont biflagellates can have one long flagellum and one short, or one smooth and one hairy (bearing mastigonemes). In our case, control specificity is all the more surprising as we only considered species whose flagella exhibit no exterior morphological differences. During flagellar assembly and lengthening, tubulin subunits are shuttled from an intracellular pool by molecular motors, and added to the tips of growing flagella by a process known as intraflagellar transport [51]. This process is unlikely to result in structural heterogeneities between the individual flagella axonemes. Instead, asymmetries arising from physiological differences in basal body age [52] and associated contractile structures may be more significant. In flagellate green algae, basal body duplication is semi-conservative. The two flagella of *C. reinhardtii* are termed *cis* or *trans* according to proximity to a single eyespot: the *trans* basal body is inherited from the parental cell but the *cis* is formed anew. This distinction likely underlies their differential calcium response, important during phototaxis [11]. The octoflagellate exhibits an extreme form of this asymmetric ultrastructural patterning, where the eight flagella are again distinguishable by the age of the attached basal body and organisation of accessory structures (Figure 5a).

Asymmetries in the control architecture itself are sufficient but not necessary to template asymmetric dynamics. A radially-symmetric nerve ring innervates the 5-legged starfish [53], yet the organism routinely performs a breaststroke gait differentiating between two pairs of side arms and a single leading arm. It would therefore be interesting to explore whether the "special" flagellum in *P. octopus* is always the same flagellum (as uniquely identified by basal body numbering), or whether different flagella can take turns to assume the "exploratory" position (SM Video 10). Given the dual sensory and motor capabilities of cilia and flagella, this could signify an early evolution of division of labour. We suggest that symmetry-breaking in excitatory or inhibitory signalling in the algal flagella apparatus specifies the local state of contractility of intracellular fibres to control the activation state of individual flagella. This may be likened to the case of vertebrate limbs, where antagonistic control of flexor and extensor muscles is provided by motor neurons, or to marine invertebrate larvae, where specialised ciliomotor neurons induce coordinated ciliary arrest [54]. We showed that in multiflagellates the ability to activate subsets of flagella provides a novel mechanism for trajectory reorientations which is distinct from steering in sperm (Alvarez *et al*, this Issue) and in other uniflagellates which rely upon a change in the beat pattern of a single flagellum. A quadriflagellate can turn left, right, up or down, depending on which of its four flagella is active. However, the extent to which organisms make use of these capabilities warrants further study.

The diverse gaits of multiflagellates may have arisen from different evolutionary pressures associated with a need to occupy different ecological niches. Even small genetic changes could cause sufficient genetic rewiring of the basal apparatus and differences in locomotor pattern. With increasing flagella number, we suggest that new gaits arose from superpositions of gaits from flagellates with fewer flagella. In this perspective of gait modularity, the octoflagellate rotary breaststroke may have originated from intercalation of two quadriflagellate trotting gaits, and the trot or gallop from intercalcation of two pairs of biflagellate gaits. Indeed, it is thought that *P. octopus* branched from a related quadriflagellate *Pyramimonas* species via incomplete cell division [48].

In conclusion, the capacity to generate and coordinate complex behaviour is by no means exclusive to organisms possessing neural circuitry. In diverse unicellular flagellates, phase-resetting processes and transitions in locomotor patterning occurred much faster than is possible by any passive means, implicating excitable signalling. In future work, we will seek to identify structures capable of functioning as pacemakers for locomotor patterning in flagellates. By analogy with CPGs, we suggest that parts of the algal cytoskeleton may be independently capable of generating coupled oscillations. This has enabled these organisms to dynamically reconfigure their gaits in response to environmental changes and uncertainty, without the need for a central controller. The latter feat harkens to the embodiment perspective applicable to designing bioinspired robots, in which "control of the whole", is "outsourced to the parts" [56]. As modern technology strives towards greater automation, engineering adaptability into artificial systems has remained a formidable challenge. In this sense, cell motility is in fact a form of physical embodiment, wherein the compliant cytoskeleton and appendages enact morphological computation. In light of these findings, we may wish to extend the scope of gait research beyond model vertebrates and invertebrates [23, 55] to include aneural organisms. Much can be gained from exploring how simple unicellular organisms sculpt motor output and achieve sensorimotor integration – for herein lies the evolutionary origins of decentralised motion control.

**Acknowledgements**
I thank Ray Goldstein and Gáspár Jékely for discussions, and an anonymous referee for suggesting the relevance of embodiment concepts from control theory. Financial support is gratefully acknowledged from the University of Exeter.

# Legends and captions for electronic supplementary materials.

**The following multimedia files (V1-10) accompany this study – all scalebars are 10 μm.**

1. V1 - A quadriflagellate gait transition from a spinning gait to a trotting gait.

   This movie shows a free-swimming *Tetraselmis subcordiformis* cell undergoing a gait transition (track corresponds to Figure 2b) from a spinning gait of two flagella, to a trotting gait of four flagella.

2. V2 - A quadriflagellate symmetry breaking gait.

   This movie shows a free-swimming *Tetraselmis subcordiformis* cell exhibiting a symmetry-breaking gait (track corresponds to Figure 2c) in which beating is restricted to only one of its four flagella.

3. V3 - Another quadriflagellate symmetry breaking gait.

   This movie shows a free-swimming *Pyramimonas parkeae* cell exhibiting a symmetry-breaking gait (track corresponds to Figure 2f) in which beating is restricted to only one of its four flagella.

4. V4 - A quadriflagellate resetting its forward-swimming gait after a shock response.

   This movie shows a free-swimming *Carteria crucifera* cell before, during, and after a spontaneous shock response (Figure 3a). Note the rapid phase-resetting dynamics of the flagella.

5. V5 - A quadriflagellate gait with two of four flagella active.

   This movie shows a free-swimming *Pyramimonas tychotreta* cell in which beating is restricted to two of its four flagella (Figure 3b).

6. V6 – A quadriflagellate being caught by micropipette aspiration.

   This movie shows a *Pyramimonas parkeae* cell with initially quiescent flagella being captured by a micropipette (Figure 4a). Flagella beating is initiated at the moment of capture.

7. V7 – Demonstrating gait-mechanosensitivity in a micropipette-fixed quadriflagellate.

   This movie shows a micropipette-fixed *Tetraselmis suecica* cell exhibiting spontaneous and induced gait transitions (Figure 4b).

8. V8 – The axial rotation of an octoflagellate during swimming.

   This movie shows the clockwise axial rotation of *Pyramimonas octopus* during free-swimming, and the synchrony between pairs of diametrically opposite flagella.

9. V9 – The rotary breaststroke of an octoflagellate during swimming.

   This movie shows the rotary breaststroke of *Pyramimonas octopus* during free-swimming (Figure 5b). The cell is viewed from posterior end.

10. V10 – The octoflagellate search gait in which one flagellum is extended.

    This movie shows the search gait of *Pyramimonas octopus* in which one flagellum is held extended in front of the cell (Figure 5e).